\documentclass[prl,superscriptaddress,showpacs,twocolumn,preprintnumbers]{revtex4}
\usepackage{graphicx}

\def\be{\begin{eqnarray}}
\def\ee{\end{eqnarray}}

\newcommand{\singlet}{{\underline{\bf 0}}}
\newcommand{\octet}{{\underline{\bf 8}}}

\newcommand{\nn}{{\nonumber}}
\newcommand{\MeV}{\mbox{\rm\,MeV}}
\newcommand{\GeV}{\mbox{\rm\,GeV}}

\newcommand{\al}{\alpha}

\newcommand{\als}{\alpha_s}
\newcommand{\ice}[1]{\relax}
\newcommand{\ba}{\begin{eqnarray}}
\newcommand{\ea}{\end{eqnarray}}

\newcommand\VV{\setbox0=\hbox{V}\hbox{\rm V\raise\ht0
  \hbox to0pt{\hss\vbox to0pt{\hbox{v}\vss}}}}
\def\slashchar#1{\setbox0=\hbox{$#1$}           % set a box for #1
   \dimen0=\wd0                                 % and get its size
   \setbox1=\hbox{/} \dimen1=\wd1               % get size of /
   \ifdim\dimen0>\dimen1                        % #1 is bigger
      \rlap{\hbox to \dimen0{\hfil/\hfil}}      % so center / in box
      #1                                        % and print #1
   \else                                        % / is bigger
      \rlap{\hbox to \dimen1{\hfil$#1$\hfil}}   % so center #1
      /                                         % and print /
   \fi}                                         %
\begin{document}

\title{Chiral condensates in massless QCD and the $U(1)_A$ boson mass}

\author{Thomas Mannel}
\author{Alexei A. Pivovarov}

\affiliation{Theoretische Elementarteilchenphysik,
  Naturwiss.- techn. Fakult\"at,
  Universit\"at Siegen, 57068 Siegen, Germany}

\begin{abstract}
\noindent The $U(1)_A$ boson mass is calculated through the phenomenological 
characteristics of the vacuum related to spontaneous breaking of
chiral symmetry in QCD. 
The mass is determined by
the mixed quark-gluon condensate 
$\langle \bar q G q \rangle$ that emerges in an appropriate
correlation function due to triangle anomaly.
For three flavor QCD in chiral limit
we find the numerical value for the $U(1)_A$ mass to be
$M_0=310\pm 50~\MeV$.
\end{abstract}

\pacs{13.35.Bv, 14.60.Ef}

\preprint{SI-HEP-2019-12}
\preprint{SFB-257-P3H-19-028}

\maketitle
The Lagrangian of QCD~\cite{qcd} with $N_f$ massless quarks is invariant 
under the chiral symmetry group 
$U(N_f)_L\otimes U(N_f)_R$~\cite{qcdch}.
The low energy spectrum of hadronic states suggests that the chiral  
symmetry group is spontaneously broken down to its vector part $U(N_f)_V$
with $V=L+R$.
Therefore one expects that there exist $N_f^2$ massless
Goldstone bosons corresponding to 
the generators of the broken subgroup $U(N_f)_A$ with
$A=L-R$~\cite{pagels}. 
The massless approximation for the $u,d$ quarks is very accurate in the real
world as their masses $m_u+m_d\sim 15~\MeV$
are significantly smaller than the infrared strong interaction
scale $\Lambda_{\rm QCD}\sim 500~\MeV$. For the strange quark with the
mass $m_s\sim 150~\MeV$ the massless approximation is still applicable
but less accurate numerically~\cite{Gasser:1984yg}.  
For $N_f=3$ 
the Goldstone modes are identified in
the observed experimental particle spectrum with the
members of the pseudoscalar octet ($\pi,K,\eta$) 
which would be
massless in the world of massless $u,d,s$ quarks.
The light excitation corresponding to 
the singlet axial current of the $U(1)_A$ subgroup
is, however, not seen  
as the most probable candidate -- the $\eta'$ 
boson -- seems to be too heavy, $m_{\eta'}=958~\MeV$.
The absence of the ninth light Goldstone-like mode in the mass spectrum 
is refereed to as the $U(1)$ problem.

The problem 
is solved by the observation that 
the singlet axial current corresponding to the $U(1)_A$ subgroup
does not conserve due to the triangle anomaly~\cite{adleran,belljack}
and, therefore, does 
not necessarily generate a massless state.
This solution is generically nonperturbative and requires a thorough
microscopic analysis of the QCD ground state~\cite{instant}. 
The mass of the $\eta'$-boson is quite large in a real world
and will contain not only contributions
related to the spontaneous breaking of chiral 
symmetry but also contributions
from the explicit breaking by the light-quark masses.    
The effect of this explicit
breaking can be significant since the mass of strange quark is not much
smaller than the QCD infrared scale.

In fact, the quantitative
discrepancy between the experiment and the chiral limit
approximation for
$N_f=3$ 
has been noticed already in the
paper on the triangle anomaly where the decay width
$\eta\to \gamma\gamma$ was discussed~\cite{adleran}.

While the chiral symmetry breaking in massless QCD has been
extensively studied at the microscopic level (e.g.~\cite{Adler:1984ri})
and is believed to be determined eventually by
instantons~\cite{tHooft:1986ooh},
the breaking effects are parametrized  at the phenomenological level
by
some chiral symmetry violating vacuum condensates.
In the present paper we discuss the mass of a $U(1)_A$ boson
in terms of phenomenological
characteristics of the QCD vacuum related to the spontaneous breaking
of chiral symmetry.
To this end, we consider the spectral functions of
some singlet and octet current correlators
and pin down the key difference between these two cases
which can be interpreted as
the generation of a mass in the singlet channel.

The octet axial current is
$
\label{nonsinglet}
j_{5\mu}^a = \bar \psi \gamma_\mu \gamma_5 (\lambda^a/2) \psi$,
where $\lambda^a$, $a=1,\ldots,8$,
are the Gell-Mann matrices for $SU(3)$,
and $\psi= (u,d,s)$.
These currents  are
conserved in the massless limit 
$
\partial^\mu j_{5\mu}^a=0
$
and generate the $SU(3)_A$ subgroup of the chiral group.
The coupling of these currents to Goldstone states
$|{\rm P}^a({\bf p})\rangle$ 
is given by the constant $f_{\octet}$, 
$
\langle 0| j_{5\mu}^a(0)|{\rm P}^b({\bf p})\rangle
=ip_\mu \delta^{ab} f_\octet
$.
The bilinear quark densities
$j_{5}^a = \bar \psi i\gamma_5 (\lambda^a/2) \psi$
also interpolate
these Goldstone states with the coupling $g_\octet$,
$
\langle 0| j_{5}^a(0)|{\rm P}^b({\bf p})\rangle
=\delta^{ab} g_\octet
$.
The relation between 
the constants $f_\octet$ and $g_\octet$ can be obtained
by introducing a small 
explicit breaking of chiral symmetry by a quark mass
term~\cite{Glashow:1967rx,phenlagr}
which we introduce into the QCD Lagrangian by a
diagonal mass matrix ${\rm diag}(m,m,m)$ that
retains the $SU(3)_V$ flavor symmetry. This leads to 
$
\partial^\mu j_{5\mu}^a= 2 m j_{5}^a
$
and one obtains a relation between 
the constants $f_\octet$ and $g_\octet$ in the form
\be
\label{eq:f8g8}
2 m g_\octet= M_P^2 f_\octet \ ,\qquad \mathrm{or}
\qquad  g_\octet=
\frac{M_P^2}{2m}
f_\octet \, .
\ee 
Here $M_P$ is a mass of the (almost)-Goldstone particles
that emerges due to an explicit chiral symmetry breaking
through the quark-mass
term in the Lagrangian.
For the case of our diagonal mass matrix
the emerging masses of the octet states are equal.
  
While the microscopic description of
the mechanism of the spontaneous breaking of chiral
symmetry
provides a dynamical picture of 
the QCD ground state as the BCS explanation of superconductivity
does,
on the phenomenological level this structure 
is reflected in emergence of order parameters
in the spirit of Landau-Ginzburg model~\cite{Ginzburg:1950sr}.
In QCD the order parameters of the chiral symmetry breaking
are chiral vacuum condensates.
The microscopic dynamics is eventually parametrized
with the phenomenological exact ground-state
characteristics which are
accounted for at the computation of correlation
functions~\cite{Politzer:1976tv}.
In our paper, we take a phenomenological route. 
\vspace*{0.2cm}
\begin{figure}[ht]
\quad\includegraphics[scale=0.6]{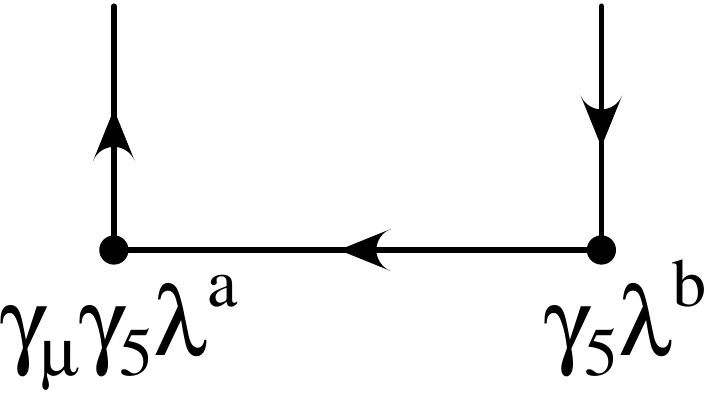}
\caption{Octet correlator}
\label{fig:octet}
\end{figure}

We start with the octet channel and consider a correlator of the form
\[ 
\label{nonsinglcorrtheor0}
\int dx e^{ipx}
\langle T j_{5\mu}^a(x) j_5^b (0)\rangle
=p_\mu\delta^{ab}\Pi_{\octet}(p^2)\, .
\]
In the chiral limit the operator product expansion 
(OPE) gives including the chiral symmetry
breaking effects through vacuum condensates
\be 
\label{nonsinglcorrtheor}
\Pi_{\octet}(p^2)=C^\octet_{ \bar q q }\frac{\langle \bar q q \rangle}{p^2}
+C^\octet_{ \bar q G q }\frac{\langle \bar q G q \rangle}{p^4}
+O\left(\frac{1}{p^6}\right)\, .
\ee
Here $q$ is a generic name for any light quark $u$, $d$, or $s$,
$\langle \bar q q\rangle$ is a quark condensate which is an
order parameter of
the spontaneous chiral symmetry breaking,
$\langle \bar q Gq\rangle$ is a mixed quark-gluon condensate, 
$\langle \bar q Gq\rangle
= \langle \bar q g_s G_{\mu\nu}^a t^a\sigma^{\mu\nu} q\rangle$,
$t^a$ are the generators of the QCD color group
$SU(N_c)$ with normalization condition
${\rm tr} (t^a t^b) =\delta^{ab}/2$, 
and $\sigma_{\mu\nu}=i[\gamma_\mu,\gamma_\nu]/2$. 

In the chiral limit, $C^\octet_{ \bar q q }=1$,
and there are neither higher order corrections
in $\als$ nor power corrections $1/p^n$ with $n>2$
(in particular, $C^\octet_{ \bar q G q }=0$) and thus 
$$
\Pi_{\octet}(p^2)=
\frac{\langle \bar q q \rangle}{p^2} \quad  \mbox{in the chiral limit}
\, .
$$   
These properties of the correlator~(\ref{nonsinglcorrtheor})
can be derived from current algebra and the canonical field 
commutators (e.g.~\cite{Treiman:1986ep}). 
They can be also established by direct computation in perturbation
theory (e.g.~\cite{Pivovarov:1992qu}) that gives a perturbative
confirmation
of canonical commutators structure
of QCD.

The Goldstone modes
give the only contribution to the correlator possible
in the chiral limit $M_P^2\to 0$
\be 
\label{nonsinglcorrexp}
\Pi_{\octet}^{\rm ph}(p^2)
=\frac{f_\octet g_\octet}{M_P^2-p^2}
\ee
Comparing eqs.~(\ref{nonsinglcorrtheor},\ref{nonsinglcorrexp}) 
one obtains 
$
\label{generalGMOR}
f_\octet g_\octet = - \langle \bar q q \rangle
$.
The Gell-Mann-Oakes-Renner
relation~\cite{Gell-Mann:rz,Politzer:1976tv}
$
\label{GMOR}
f_\octet^2 M_P^2 = - 2 m   \langle \bar q q \rangle\, 
$ 
follows
using eq.~(\ref{eq:f8g8}).

\vspace*{0.2cm}
\begin{figure}[t]
\quad\includegraphics[scale=0.4]{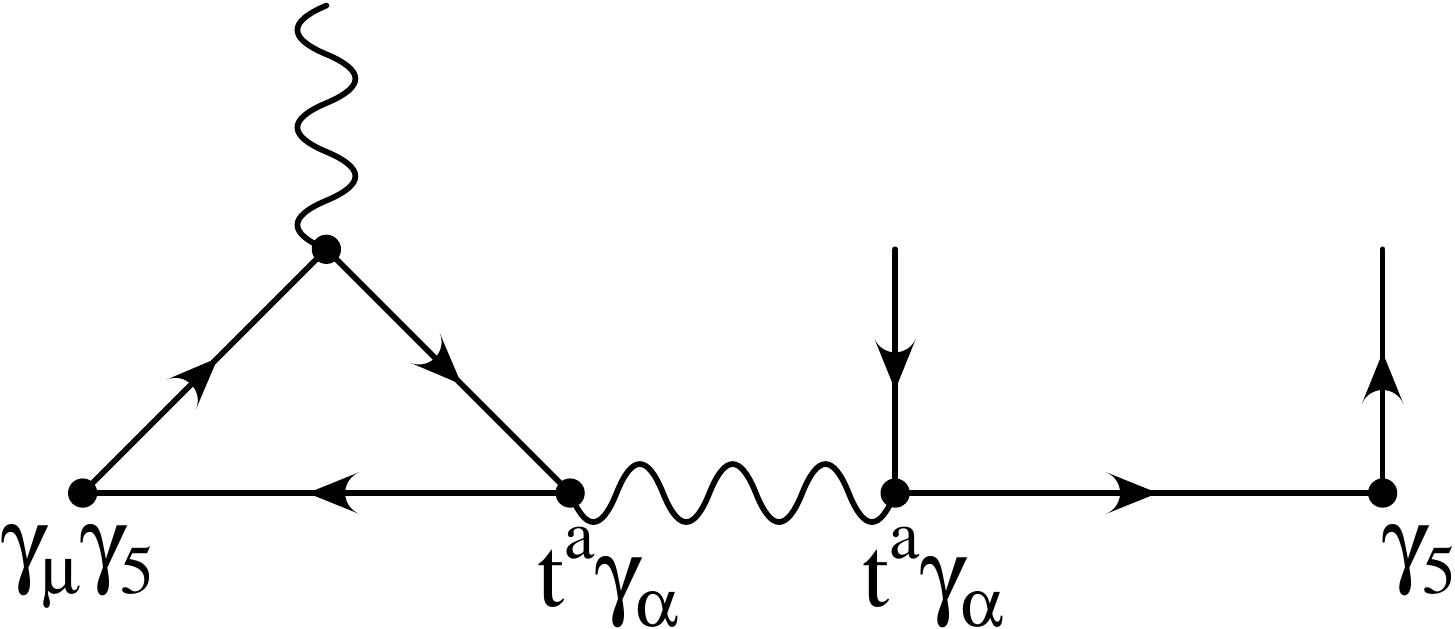}
\caption{Triangle diagram contribution to the singlet correlator}
\label{fig:fig2}
\end{figure}

For the correlator of the singlet currents 
the difference in the OPE 
is a presence of a triangle diagram
as given in Fig.~\ref{fig:fig2}.
The anomaly~\cite{adleran,belljack} gives contributions of higher order in $1/p^2$
and, in principle, in higher orders in $\als$ for the leading quark
condensate
term. One has
\ba 
\label{singcorr}
\int dx e^{ipx}
\langle T j_{5\mu}(x)  j_{5}(0)\rangle
=p_\mu\Pi_\singlet(p^2)\, , \\
\label{singlcorrtheor}
\Pi_\singlet(p^2)=C_{ \bar q q }^\singlet \frac{\langle \bar q q \rangle}{p^2}
+C_{ \bar q G q }^\singlet \frac{\langle \bar q G q \rangle}{p^4}
+O\left(\frac{1}{p^6}\right)
\ea
where the normalization $\lambda^0/2={\rm diag}(1,1,1)/\sqrt{6}$
has been used for singlet operators,
$j_{5\mu}= \bar \psi \gamma_\mu \gamma_5 (\lambda^0/2) \psi$.
In the chiral limit and to the leading order in $\als$ one finds 
$C_{ \bar q q }^\singlet =1+O(\als^2)$ and
$C_{ \bar q G q }^\singlet =3\als/4\pi+O(\als^2)$.
Higher order corrections to the expansion of the 
correlator in both $\als$ and $1/p^2$ are not forbidden.  
Thus, in the framework of
operator product expansion with phenomenological account for
spontaneous symmetry breaking
one clearly sees the difference between
singlet and octet channels in the chiral limit.

We analyze the experimental spectrum within the framework of
spectral sum rules~\cite{Glashow:1967iv,Poggio:1975af,Shankar:1977ap}.
We use
\be 
\label{sumrules}
\oint_C\Pi^{\rm ph}(z)z^n dz=\oint_C\Pi(z)z^n dz
\ee
for $n=1,2$~\cite{fesr}
and analyze
the contribution of the $U(1)_A$ boson to the
correlator~(\ref{singcorr})
\be 
\label{singph}
\Pi^{\rm ph}_\singlet (p^2)=\frac{f_\singlet g_\singlet }{M_0^2-p^2}
\ee
with 
$\langle 0| j_{5\mu}(0)|{\rm P}({\bf p})\rangle=ip_\mu f_\singlet $ and 
$\langle 0| j_5 (0) |{\rm P}({\bf p})\rangle = g_\singlet $.
Choosing an integration contour that encircles only the lowest state
and comparing expressions (\ref{singlcorrtheor}) and
(\ref{singph}) one finds the first sum rule ($n=1$) 
\be
f_\singlet g_\singlet  = - C_{\bar q q}^\singlet
\langle \bar q q \rangle\, .
\ee 
The second sum rule ($n=2$) gives 
\be
f_\singlet  g_\singlet  M_0^2 = - C_{\bar q Gq}^\singlet
\langle \bar q Gq \rangle
\ee
which leads to an expression for the $U(1)_A$ boson mass
in the chiral limit
\be 
\label{eq:u1-mass}
M_0^2=\frac{C_{\bar q Gq}^\singlet }{C_{\bar q q}^\singlet }
\frac{\langle \bar q Gq \rangle}{\langle \bar q q \rangle}\, .
\ee
Using the parametrization
$\langle \bar q Gq \rangle = 
m_0^2 \langle \bar q q \rangle$~\cite{Belyaev:sa} 
and explicit expressions for the coefficient functions one finds 
(in the chiral limit and to the leading order in $\als$)
\be 
M_0^2=\frac{3\al_s}{4\pi}m_0^2\left(1+O(\als)\right)+O(m_s)\ .
\ee
Numerically, with $m_0^2=0.8\pm 0.2\GeV^2$~\cite{m0}
and $\al_s(1\GeV)=0.50\pm 0.05$~\cite{als}
one gets 
$M_0^2=0.10\pm 0.04~\GeV^2$ and $M_0=310\pm 50~\MeV$.
The usual candidate for the $U(1)_A$ boson in the
observed particle spectrum
is the $\eta'$ boson with $m_{\eta'}=958\MeV$.
Thus we conclude that the 
anomaly related contribution to the mass of $\eta^\prime$ in the chiral limit
is only about 1/3 of the observed
${\eta'}$-mass.

Corrections to this estimate come from
higher orders in the OPE and from the explicit chiral symmetry breaking 
by quark masses. The perturbative  
corrections to the coefficient
functions of condensates are calculable due to asymptotic freedom of 
QCD~\cite{Gross:1973id} but can be significant as the expansion parameter 
$\al_s(1\GeV)$ is large (e.g.~\cite{glu,Gabadadze:1994tn}). 
The largest part of the corrections from explicit breaking of chiral symmetry through 
the quark masses comes from the strange quark, since $m_s$ is sizable and 
leads to a breaking of the flavor $SU(3)$ symmetry by the mass itself and 
by the different numerical values of $\langle \bar{s} s \rangle$ compared to the 
light-quark condensates~(e.g.~\cite{Gimenez:vg}).

There are also corrections on the hadronic side
which come from considering more complicated spectra including 
mixing with gluonia and the $\eta$ state from the pseudoscalar octet.
In this case higher dimensional condensates with dimensionality larger
than five in the OPE eq.~(\ref{singlcorrtheor}) can play
some role.

We consider the $s$-quark mass $m_s$ corrections as the most essential
ones for quantitative comparison with experiment.
The mass matrix of the form $m_s{\rm diag}(0,0,1)$
leads at leading order in $m_s$ 
to the corrections to the OPE
\be 
\label{eq:mscorrnondiag}
\Pi_{m}^\singlet(p^2)= \frac{m_s}{6}\Delta(p^2), \quad
\Pi_{m}^{\octet\singlet}(p^2)=
-\frac{m_s}{3\sqrt{2}}\Delta(p^2) 
\ee
for the singlet and singlet-octet correlators, 
and to 
\be 
\Pi_{m}^K(p^2)= \frac{m_s}{4}\Delta(p^2), \quad
\Pi_{m}^\eta(p^2)= \frac{m_s}{3}\Delta(p^2) 
\ee
for the octet correlator.
The leading order
perturbative expression for the spectrum of $\Delta(p^2)$
reads
\be
\Delta(s)=\frac{3}{4\pi^2}
\left(1+O(\als(\mu))\right)\, .
\ee
The corrections in $\als(\mu)$ can be large
at the relevant scale $\mu\sim 1~\GeV$ (e.g.~\cite{msgammaovch}).
Furthermore we neglect the term $m_s\langle GG\rangle $
where $\langle GG\rangle$ is a gluon condensate~\cite{Shifman:bx}.
The $m_s$ octet corrections obey the symmetry relation
$ 
4\Pi_{m}^K(p^2)= 3\Pi_{m}^\eta(p^2)+ \Pi_{m}^\pi(p^2)
$
with $\Pi_{m}^\pi(p^2)=0$ and reproduce the octet mass formula
$3 m_\eta^2+m_\pi^2=4m_K^2$.

The mass shift to the chiral-limit prediction~(\ref{eq:u1-mass})
depends on details of the spectrum.
The mass of the $s$-quark is not negligible compared
to $\Lambda_{\rm QCD}$ which manifests itself in the fact that 
both the kaon and the $\eta$ are rather massive in the real world.

The explicit breaking of the flavor symmetry induces
an octet-singlet mixing. Even if the violation term is small 
the mixing effect can be
large.
Thus, for vector mesons the mixing is ideal --
the $\phi$ meson is almost exclusively an $\bar s s$ state
while the $\omega$ meson is built from $u,d$ quarks only. 
The $\eta,\eta'$ mixing has been extensively studied
(e.g.~\cite{Feldmann:1998vh}). 
For our purposes we consider contributions of $\eta$ and $\eta'$
to the correlators
\ba
\label{eq:fullPh}
&&\Pi^{ph}_\octet(p^2) =\frac{f_\octet g_\octet}{M_\eta^2-p^2} + 
\frac{f_{\octet\singlet} g_{\octet\singlet}}{M_0^2-p^2} \nn \\
&&\label{eq:singletms}\Pi^{ph}_\singlet(p^2)
=\frac{f_\singlet g_\singlet }{M_0^2-p^2}+
\frac{f_{\singlet\octet} g_{\singlet\octet}}{M_\eta^2-p^2} \\
&&\label{eq:octetsinglems}
\Pi^{ph}_{\octet\singlet}(p^2)  =
\frac{f_{\octet} g_{\singlet\octet}}{M_\eta^2-p^2} + 
\frac{f_{\octet\singlet} g_{\singlet}}{M_0^2-p^2} 
\ea
Here $f_{\octet\singlet }$ is a non-diagonal overlap of the
octet current with $\eta'$,
$
\label{eq:matr-el}
f_{\octet\singlet }\sim \langle 0|j_\mu^a|\eta' \rangle$,
same for $g_{\octet\singlet}$, while 
$f_{\singlet \octet }\sim \langle 0|j_5^\mu|\eta \rangle$ and similar
for 
$g_{\singlet \octet}$.
The diagonal constants do not vanish in the symmetry limit and acquire
a correction in the symmetry breaking parameter $m_s$,
e.g. $f_{\octet} \to f_{\octet}+O(m_s)$. The nondiagonal
matrix elements vanish in
the symmetry limit, $f_{80}=O(m_s)$, as one can conclude
from
eq.~(\ref{eq:octetsinglems}) using eq.~(\ref{eq:mscorrnondiag}).
To the linear order in $m_s$ one neglects the
nondiagonal contributions in eq.~(\ref{eq:singletms})
which are of $O(m_s^2)$ order and gets
\ba
\label{mscorrSR}
&&f g =- \langle \bar q q \rangle 
+ \frac{m_s}{8\pi^2}s_0 \nn \\
&&f g M_0^2=-\langle \bar q q \rangle
\frac{3\al_s}{4\pi}m_0^2 
+\frac{m_s}{16\pi^2}s_0^2
\ea
where $f,g$ are the singlet parameters up to $O(m_s)$.
We take the numerical 
value $m_s(1\GeV)=150\MeV$~\cite{msnum}
and
$\langle \bar q q \rangle=\langle \bar u u
\rangle=-(250\MeV)^3$ from data on $f_\pi$ and
light quark masses~\cite{Gasser:1982ap}. 
One finds
$M_0^2=0.14$ for $s_0=1\GeV^2$
and 
$M_0^2=0.2$ for $s_0=1.5\GeV^2$, or $M_0=460\MeV$.
The shift amounts to about $0.1\GeV^2$ that is comparable to the
shift in the octet, $m_K^2=0.25\GeV^2$, $m_\eta^2=0.3\GeV^2$.
Eventually the duality interval, i.e.
the value of $s_0$ can be fixed from the shifts in the
octet channel or by some other method of sum rules analysis
but presently
we prefer to choose it just in the region of $(1-1.5)\GeV^2$
for the rough estimates.
The duality intervals in singlet and octet channels differ for
nonvanishing $m_s$ that also introduces an uncertainty in $s_0$ 
determination. 
One sees that the $m_s$ corrections are numerically comparable with the 
anomaly contribution. 
Indeed, for $s_0=1.5\GeV^2$
the contributions of the anomaly and of explicit breaking corrections
are distributed according 
to $0.015|_{an}+0.003|_{ms}$ for the first sum rule and 
$ 0.0015|_{an}+0.0021|_{ms}$ for the second.
We checked that changes due to the gluon condensate emerging as
a power correction in $\Delta(s)$ is well within uncertainties
of our simple estimates. 
Thus, in leading order approximation and unless the strong coupling
corrections are very large, the $U(1)$ boson mass
lies in the range $400-600~\MeV$ that is smaller
than the $\eta'$ mass of $1\GeV$.
We have implicitly accounted for possible strong coupling
corrections to $\Delta(s)$ and $C_{\bar q Gq}^\singlet $ by taking a very
broad range of numerical values for parameters $m_s$, $m_0^2$ and
$\alpha_s$.
This range can look a bit unusual for the experts and introduces large
uncertainties in quantitative estimates but we believe
it helps to keep our qualitative conclusions stable against inclusion
of higher order corrections.
While the ratio of the condensates
$\langle {\bar q} Gq\rangle/\langle {\bar q}q\rangle$
is of a natural scale
in terms of $\Lambda_{rm QCD}$,
the numerical smallness of the mass is due to
an explicit suppression by $\als$.
The result for $U(1)_A$ boson mass scales as $1/N_c$ in the large
$N_c$ limit that also suggests the smallness of the numerical value.
The shift due to an explicit breaking with $m_s$ is significant:
it is not suppressed by $\als$, it is not suppressed by $1/N_c$, and
$m_s$ is not drastically small compared to $\Lambda_{\rm QCD}$. 
Still the total result seems to be small to hit the $\eta'$ mass.
And the shift due to the explicit breaking cannot
be arbitrarily large as it is
constrained from the octet channel. 

Can it happen that the proper $U(1)$ boson is not discovered yet?
In the region $450-600\MeV$
there is a broad bump $f_0(500)$
which is identified with a scalar $\sigma$~\cite{Pelaez:2015qba}.
Given the fact that our prediction for the $U(1)$ boson mass
is in this region, one could speculate about 
an isotopic singlet pseudoscalar $\eta_1$
hiding there.
If such a state would be found it could be interpreted as a $U(1)$
boson instead of $\eta'$.
Such a state could be handy as it might be a building block for the
scalar states in the molecular picture of massive scalars,
$a_0(980)\sim \eta_1\pi$. 

The $U(1)$-boson as a singlet can be strongly coupled
to the gluon sector of QCD
which is expressed by
the Witten-Veneziano formula~\cite{Witten:1979vv,Veneziano:1979ec}
$
m_{\eta'}^2=12\chi_{top}/f_{\eta'}^2
$
where 
\[
\chi_{top}=\left(\frac{\alpha_s}{4\pi}\right)^2
\int dx\langle F\tilde F(x) F\tilde F(0) \rangle\, .
\]
From our result~(\ref{eq:u1-mass}) one can extract the numerical value
for the vacuum susceptibility in the chiral limit.

To summarize,
we have calculated the mass of the particle associated with the $U(1)_A$
generator of chiral symmetry in QCD. This particle
acquires a nonvanishing mass because of the singlet axial anomaly.
We use phenomenological
characteristics of the vacuum implicitly
reflecting the nonperturbative structure of QCD in
order to compute the mass.
The obtained value is, however, too
small to be associated with the $\eta'$  
state in observed spectrum.
Even taking into account a nonvanishing $s$-quark mass is still
insufficient to close this gap so we speculate 
about the possibility that the proper $U(1)$ particle is still hidden in
the debris of a big bump around $f_0(500)$.

\begin{acknowledgments}
This work 
is supported by the Deutsche Forschungsgemeinschaft 
(DFG, German Research Foundation) under grant  396021762 - TRR 257 
``Particle Physics Phenomenology after the Higgs Discovery'' .
\end{acknowledgments}

\end{document}